\documentclass[10pt,twocolumn,letterpaper]{article}

\usepackage[pagenumbers]{wacv} 

\usepackage[accsupp]{axessibility}  
\usepackage{graphicx}
\usepackage{amsmath}
\usepackage{amssymb}
\usepackage{booktabs}

\usepackage[pagebackref,breaklinks,colorlinks]{hyperref}
\usepackage{multirow}
\usepackage{float}

\usepackage[capitalize]{cleveref}
\crefname{section}{Sec.}{Secs.}
\Crefname{section}{Section}{Sections}
\Crefname{table}{Table}{Tables}
\crefname{table}{Tab.}{Tabs.}

\def\wacvPaperID{2226} 
\def\confName{WACV}
\def\confYear{2024}

\begin{document}
\title{VCISR: Blind Single Image Super-Resolution \\
with Video Compression Synthetic Data}

\author{Boyang Wang{$^*\dagger$}, Bowen Liu{$^*$}, Shiyu Liu{$^*$}, Fengyu Yang\\
University of Michigan, Ann Arbor\\
{\tt\small \{boyangwa, bowenliu, shiyuliu, fredyang\}@umich.edu}
}
\maketitle
\def\thefootnote{*}\footnotetext{Authors contributed equally to this work.}
\def\thefootnote{$\dagger$}\footnotetext{Corresponding author.}
\begin{abstract}
   In the blind single image super-resolution (SISR) task, existing works have been successful in restoring image-level unknown degradations. However, when a single video frame becomes the input, these works usually fail to address degradations caused by video compression, such as mosquito noise, ringing, blockiness, and staircase noise. In this work, we for the first time, present a video compression-based degradation model to synthesize low-resolution image data in the blind SISR task. Our proposed image synthesizing method is widely applicable to existing image datasets, so that a single degraded image can contain distortions caused by the lossy video compression algorithms. This overcomes the leak of feature diversity in video data and thus retains the training efficiency. By introducing video coding artifacts to SISR degradation models, neural networks can super-resolve images with the ability to restore video compression degradations, and achieve better results on restoring generic distortions caused by image compression as well. Our proposed approach achieves superior performance in SOTA no-reference Image Quality Assessment, and shows better visual quality on various datasets. In addition, we evaluate the SISR neural network trained with our degradation model on video super-resolution (VSR) datasets. Compared to architectures specifically designed for the VSR purpose, our method exhibits similar or better performance, evidencing that the presented strategy on infusing video-based degradation is generalizable to address more complicated compression artifacts even without temporal cues. The code is available at \href{https://github.com/Kiteretsu77/VCISR-official}{\textit{https://github.com/Kiteretsu77/VCISR-official}}.
\end{abstract}

\begin{figure*}
	\begin{center}
		\includegraphics[width=0.96\linewidth]{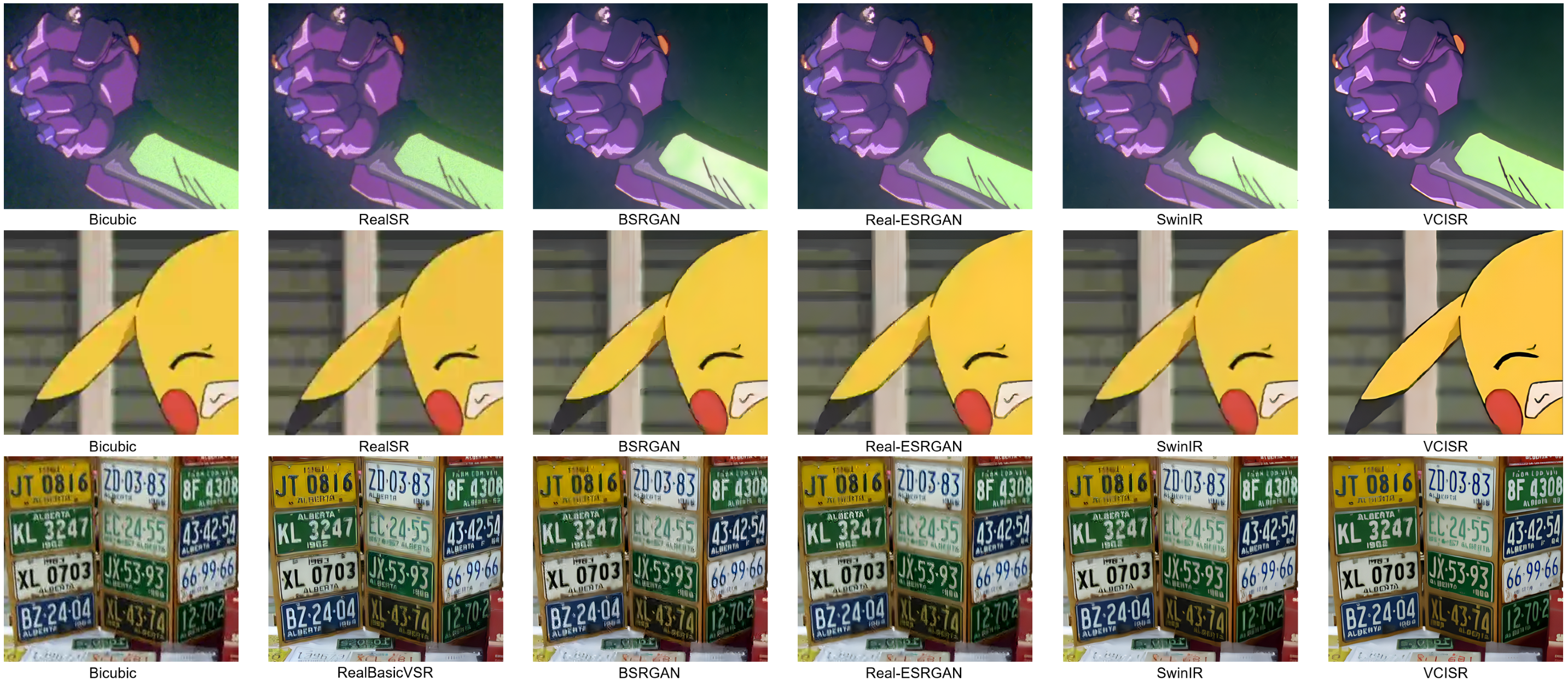}
	\end{center}
  \vspace{-10pt}
	\caption{Qualitative comparisons of the bicubic-upsampled baseline, and RealSR\cite{ji2020real}, BSRGAN\cite{zhang2021designing}, Real-ESRGAN\cite{wang2018esrgan}, SwinIR\cite{liang2021swinir}, RealBasicVSR\cite{chan2022investigating}, and our proposed VCISR super-resolved real-world images. The SR network trained with our proposed data synthesis and degradation block can produce finer details and more visually appealing results. \textbf{(Zoom in for best view.)}}
        \vspace{-5pt}
	\label{fig:teaser}
\end{figure*}

\section{Introduction}
\label{sec:intro}
Single image super-resolution (SISR) aims at reconstructing a low-resolution (LR) image into a high-resolution (HR) one. Traditional image upsampling methods include interpolation techniques such as bicubic, nearest neighbor, and bilinear, which calculate sub-pixel values from surrounding pixel values. Since SRCNN\cite{dong2015image}, super-resolution (SR) focuses on using convolution neural network techniques to generate HR images\cite{wang2018esrgan, ledig2017photo, shi2016real, lim2017enhanced, zhang2023emergence}. They first collect the HR datasets and then create paired LR datasets through bicubic downsampling. Nevertheless, the bicubic degradation method deviates from real-world degradations, making it difficult for neural networks to restore real-world LR images.

Real-world images often contain various complex degradations, such as camera out-of-focus blurring, exposure sensor noise, read noise, analog-to-digital converters noise, and lossy data compression artifacts \cite{liu2020unified, liu2021deep, liu2023mmvc}. This phenomenon raises the field of blind SISR \cite{wang2021real, Ji_2020_CVPR_Workshops, zhang2021designing, liu2013bayesian, liu2022blind, yue2022blind, wang2023exploiting}, where input LR images may contain any real-world degradation, and neural networks need to learn how to restore these artifacts while upscaling the resolution. 

Blind SISR works synthesize LR images from HR images using a degradation model. The closer the synthesized LR images are to the real-world degradations, the more effective the network can learn to generate better visual-quality HR images \cite{zhang2021designing}. Previous works\cite{wang2021real, Ji_2020_CVPR_Workshops, zhang2021designing, luo2022learning} aim at adopting blurring, noise, and image compression in their degradation synthesis. 
However, they overlook some scenarios in the real world. For example, part of the LR input content could be a video frame, and some SR implementations split a video into frames and super-resolve each frame individually as SISR. In these scenarios, it is crucial to consider the impact of video compression artifacts, temporal and spectral distortions (\eg, mosquito noise and blockiness) due to lossy data compression, on images. 
Yet simply using frames from compressed video clips for image network training \cite{wu2022animesr, chan2022investigating} can be prone to significant training costs. This is due to the fact that there are much fewer distinct features the network can learn from video datasets, where objects and scenes share great similarity between frames, compared to image datasets (\eg, DIV2K\cite{agustsson2017ntire}).
To address the aforementioned challenges, we are motivated to design a video compression degradation model to synthesize LR images with video compression artifacts for image SR networks under image training datasets. 
Meanwhile, introducing our model benefits image compression restoration in SR applications, primarily because we are not likely to have prior knowledge of the image compression algorithm used on input. It may not be the most widely-used JPEG algorithm\cite{wallace1992jpeg}. Some image compression algorithms, like WebP\cite{si2016research}, may involve intra-prediction techniques that are widely used in video compression but not in JPEG. 
Thus, we argue that video compression artifacts are capable of approximating a wider variety of distortions caused by image compression algorithms in the real world.

After reviewing video codecs and compression settings adopted in existing video SR (VSR) works\cite{chan2022investigating, wu2022animesr, li2021comisr, yang2022learned, caballero2017real}, we find that the video compression based degradation models they use are too scattered to serve as a proximity to real-world video coding artifacts. Consequently, networks trained on these scattered video compression degradations are hard to restore complicated distortions (\eg, blockiness with basis patterns), even with the aid of temporal domain propagation paths across the frames to exploit more correlations. This observation advocates us to evaluate the same network trained for blind SISR on low-quality video datasets.

Our experiments confirm that temporal compression artifacts can be simulated with spatial-only information. This allows us to synthesize video artifacts on common image SR datasets (\eg, DIV2K\cite{agustsson2017ntire}), as a manner to facilitate image SR network training. To intensify these compression artifacts in images, we present a comprehensive degradation model, which promotes the qualitative and quantitative performance of the trained SR network to a new level.

Furthermore, we propose an image dataset that contains versatile compression artifacts, which broadly exist in real-world images. This dataset is targeted to become a guideline for future researchers on how compression distortions may appear in the real world. 

Our contributions can be summarized as follows:
\begin{itemize}
    \item This work uses image super-resolution network and image training datasets to restore video and broader compression quality loss by introducing video artifacts in the degradation model. As a result, our proposed method is competitive with SR networks on real-world image and video restoration.
    
    \item
    We introduce VC-RealLQ, a real-world image dataset consisting of versatile temporal and spatial compression artifacts from various contents, resolutions, and compression algorithms. Our dataset could serve as a common benchmark for future methods and will be released for ease of future research. 
    
    \item The proposed video compression-based degradation block can be directly adopted by widely used blind SISR degradation models with minimal effort.
\end{itemize}

\begin{figure*}
	\vspace{-2mm}
	\begin{center}
		\includegraphics[width=0.9\linewidth]{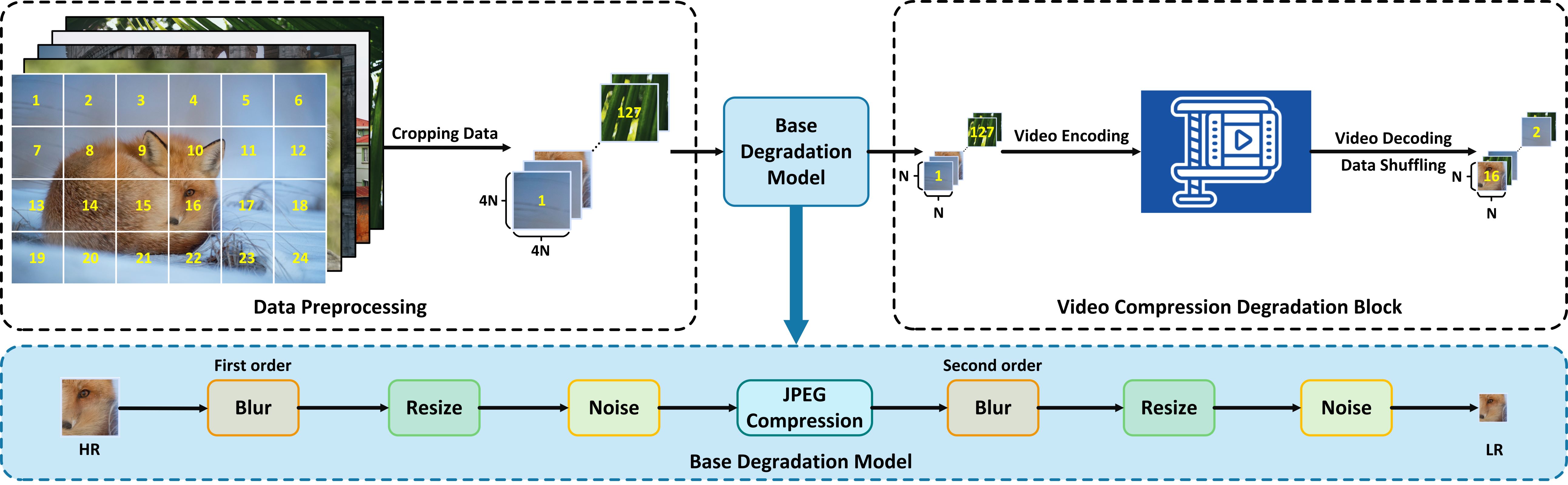}
	\end{center}
	\caption{Overview of the proposed video compression degradation model. We use the degradation model introduced in \cite{wang2021real} as a backbone of our method, and our approach comprises a data preprocessing step and a video-based degradation block.}
	\label{fig:degradation_model}
	\vspace{-4mm}
\end{figure*}

\begin{figure}[t]
    \centering
    \includegraphics[width=0.96\linewidth]{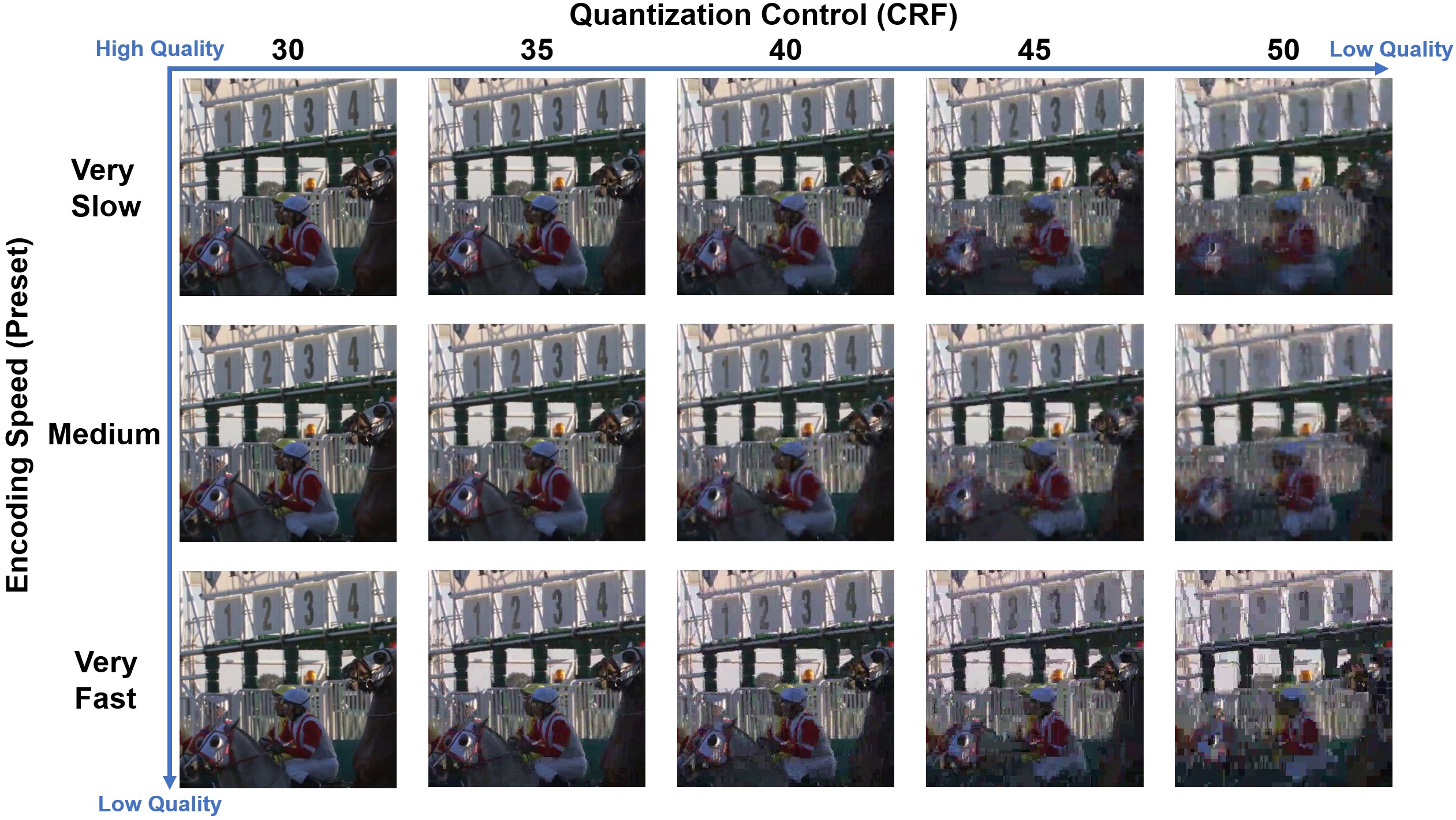}
    \caption{Encode Speed vs. Quantization Control. Under the same Quantization level (\textit{CRF}), a faster encode speed leads to more high-frequency information (margins and details of objects) lost. This phenomenon becomes more severe for a higher \textit{CRF} value. No other noise is introduced in this comparison. The frame comes from UVG\cite{mercat2020uvg} \textbf{(Zoom in for best view.)}} 
    \label{fig:CRF_preset}
    \vspace{-10pt}
\end{figure}

\section{Related Works}
\label{sec:related_works}
\subsection{Deep Blind Image SR Networks}
Blind SISR aims to upscale and restore LR images with unknown degradations. The study in this field has achieved substantial progress in recent years due to the advancement of deep neural networks~\cite{He2015DeepRL, Zhao2022RBCRT, MaQBL23, li2023interpretable, Yang2022SparseAC}. A mainstream of the existing methods adopts CNN-based~\cite{He2015DeepRL, yang2023generating, wu2022adma, chen2023search, yang2022touch} building blocks to their network architectures (\eg RealSR \cite{ji2020real}, BSRGAN\cite{zhang2021designing}, Real-ESRGAN\cite{wang2021real}). In particular, multiple levels of residual and/or dense blocks are widely involved to improve the network depth and the restoration quality. With the Transformer gaining promising performance in a variety of vision tasks\cite{dosovitskiy2020image, Wu_2023_boosting}, some latest SISR frameworks (\eg SwinIR\cite{liang2021swinir}, VRT\cite{liang2022vrt}, GRL\cite{li2023efficient}) start to include Transformer-based blocks, aiming to better capture the long-range dependencies and enhance the capacity of representation learning to facilitate image restoration and super-resolution.

\subsection{Degradation Models}
\noindent\textbf{Image degradation models.} Recent deep blind SISR networks \cite{liang2021swinir, luo2022deep, Ji_2020_CVPR_Workshops, luo2022learning, fu2022kxnet, liu2013bayesian, liang2021mutual} are mostly trained with LR images generated from the HR ones by explicit degradation model. Their models use similar degradation elements and follow a fundamental pattern:
\begin{equation}\label{equ:degradation_pattern}
  \mathbf{LR}= [\!(\mathbf{HR}\otimes \mathbf{k})\downarrow_{s} + \mathbf{n}]_{\mathtt{JPEG}.}
\end{equation}
First, HR images are convolved with kernel $\mathbf{k}$  \cite{liu2020estimating, efrat2013accurate, zhang2022deep} to simulate blurring from out-of-focus camera capture. Then, it is followed by a downsampling operation with scale factor $s$. Noises $\mathbf{n}$ are then injected into the LR images.
Finally, images will be compressed by the JPEG\cite{wallace1992jpeg} to introduce compression artifacts. 

From the aforementioned set of degradation elements (Eq. \ref{equ:degradation_pattern}), Wang~\etal \cite{wang2021real} propose a high-order degradation model which repeats the steps of blurring, resizing, adding noise, and JPEG compression a second turn. To better simulate compression artifacts, \textit{sinc} filters are adopted to create pseudo-ringing distortions.
Other works \cite{liang2021swinir, zhang2021designing, li2023efficient} employ a randomly shuffled degradation model to select degradation modules from a pool containing blur, resize, noise, and JPEG compression modules. 
After that, there are also works \cite{zhang2022closer, zhao2023quality} that combine both randomly shuffled and high-order degradation together with a skip mechanism to increase the performance. 

Overall, we observe that the existing LR image synthesis flows are still insufficient to address the intricacy of real-world images, which limits the generality and practical usage of SISR networks. Specifically, previous works only consider image-level compression artifacts, but some images may contain temporally correlated artifacts, like contents from video. To mitigate this domain gap, we propose the video compression degradation element, in the image degradation pipeline.

\noindent\textbf{Video degradation models.} 
Previous works in video SR tasks \cite{wu2022animesr, li2021comisr, khani2021efficient, yang2022learned, xiang2021boosting, chan2022investigating} mostly use H.264 \cite{schwarz2007overview} and H.265 \cite{sullivan2012overview} in their degradation model. COMSIR \cite{li2021comisr} proposes a compression-informed model for super-resolving LR videos. They only adopt H.264 in their experiment with a \textit{CRF} value (encoder parameter for QP control) between 15 and 25 for training degradation.  Khani~\etal\cite{khani2021efficient} adopt a lightweight SR model to augment H.265, They use a fixed slow mode preset to consolidate their ideas. 
On the contrary, our work uses video compression artifacts to serve as a generic degradation model for SR, not exclusively for images or videos by adopting broader \textit{preset} modes and combining \textit{preset} with \textit{QP} control to advance toward better-represented distortions. Quantization control in RealBasicVSR \cite{chan2022investigating} is enforced by selecting bitrate from a predefined range. However, bitrate only constrains the data size of the compressed video per second, and the influence of FPS (frames per second) is not considered as a result. With a higher FPS, the data budget distributed to each frame is decreased and the image frame quality may be dramatically degraded. To better simulate real-world video scenarios, we consider both bitrate and FPS influence as a combination.
To the best of our knowledge, we are the first work that enriches the blind SISR degradation model with video coding-based artifacts, which are simulated on image datasets without the loss of generality. With this technique, we can produce more realistic compression artifacts in degraded LR images to facilitate SR network training, which previous SR works have yet to investigate.

\section{Proposed Methods}
\subsection{Base Degradation Elements}
\label{sec:sr_degradation}
The blind SISR degradation approaches for LR image synthesis in existing works share a collection of modules in common as follows.

\noindent\textbf{Blur.} Blurring artifact is introduced by convolving the high-resolution with isotropic or anisotropic Gaussian blur kernels under the regular-shaped, generalized-shaped, or plateau-shaped distributions.

\noindent\textbf{Resize.} Resizing includes bicubic, bilinear, or area interpolation operations. Both downsampling and upsampling are considered to cover broader resize scenarios used in the real world.

\noindent\textbf{Noise.} The resized LR images will be added with additive synthetic noise: Gaussian noise and Poisson noise. For Gaussian noise, we perform both speckle gray noise (same synthetic noise for all RGB channels) and color noise (different synthetic noise for each RGB channel). For Poisson noise, the intensity of each pixel is independent and is proportional to its pixel intensity.

\noindent\textbf{JPEG compression.} By the end of a degradation pass, JPEG compression is introduced. It first converts RGB images into luma and chroma components (YCbCr color space). Then, each independent 8 $\times$ 8 block is self-encoded by discrete cosine transform (DCT). To synthesize compression artifacts, the DCT-transformed blocks are quantized by a quality factor $q\in[0, 100]$, where a lower $q$ indicates a lower quality. Lastly, the quantized blocks are transformed back by inverse DCT (IDCT) and converted from YCbCr color space to RGB color space.

\begin{figure*}
	\begin{center}
		\includegraphics[width=0.9\linewidth]{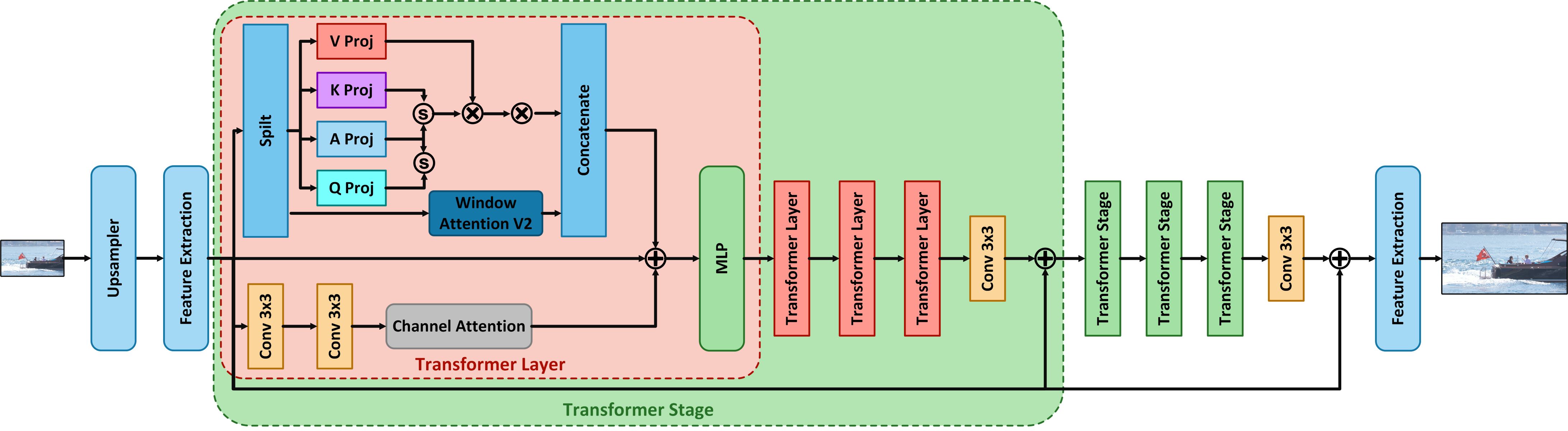}
	\end{center}
  \vspace{-10pt}
	\caption{The super-resolution network architecture. Our method adopts a smaller model of GRL\cite{li2023efficient} for x4 super-resolution. }
	\label{fig:grl_network}
 \vspace{-10pt}
\end{figure*}

\subsection{Modeling Lossy Video Compression Artifacts}
\label{sec:video_compression}
All conventional video compression standards follow the same hybrid coding algorithm. In these schemes, there are a very limited number of macroblocks
(a compression processing unit usually with sizes of 4x4, 8x8, 16x16, or even larger for some compression standards)
being self-encoded, and searching algorithms aid most macroblocks to correlate with a reference that has the closest pixel content similarity. 
The reference may come from the current frame (intra-prediction) or frames before or after it (inter-prediction). 
Moreover, the search process is constrained by the limited computation resources to compare with every block in the scope. Hence, the resulting reference blocks may only be sub-optimal in most cases. 

This mechanism is not used by JPEG, which only incorporates self-encoding within fixed-size macroblocks without finding any reference. Our proposed degradation block is therefore inclusive of a wider variety of real-world codecs, including JPEG.

With a reference in hand, a pixel-level residual between the target and reference macroblock is calculated. This residual will be converted into the frequency domain by Discrete Cosine Transform \cite{ahmed1974discrete} or Wavelet Transform \cite{yang2022ntire}. Each value of the transformed matrix will be divided by a common quantization parameter (\textit{QP}) to increase the compression rate. However, the higher the \textit{QP}, the more irreversible high-frequency information on the sources will be lost. Perceptually, the borders of the object become vague, and more blocking and noise artifacts may occur. Note that the purpose of macroblock reference prediction is to minimize the residual information as small as possible, such that it will be less influenced by the \textit{QP}. Thus, under the same \textit{QP}, macroblocks with better prediction schemes will present better visual quality as demonstrated in Figure \ref{fig:CRF_preset}.

     In summary, the quality of the compressed video is affected by 
     \textbf{1)} quantization intensity and 
     \textbf{2)} how well the searching algorithm can predict references for macroblocks in a limited preset time.

\subsection{Video Compression Degradation Model}
\label{sec:video_artifact_degradation_block}
The video compression degradation model includes a dataset preparation stage and video compression module as shown in Figure \ref{fig:degradation_model}.

\noindent\textbf{Dataset preprocessing.}
In the dataset preparation stage, images are cropped to non-overlap patches and aligned from left to right then top to bottom. We desire to synthesize temporal compression artifacts from spatially-correlated patches. Our insight comes from the understanding of compression algorithms. In compression, lossy contents (artifacts) come from the quantization of residual information between the ground truth source and the most similar reference selected by the algorithm. 
In regular video compression, the codec algorithm needs to find either a spatially-correlated reference from the same frame or a temporally-correlated reference from the nearby frames. In this context, when spatially-correlated patches become temporally-correlated patches, the spatial reference searching functionality of the codec algorithm is now turned to temporal searching. This change does not affect the fundamental methodology of compression but helps us to create more generic and complex compression artifacts to heuristically simulate distortions in the real world.

To best utilize the gap between the selected cropped patches and the full image height and width, we apply random padding on the left or the top of the image to contain more image zones as a form of augmentation.
The degradation batch size is selected to be 128, which is an empirical value that can provide a large enough window for patches to run video compression with the effectiveness of the video encoder parameters (\eg, \textit{preset}, \textit{CRF}). To avoid the keyframe selection bias, we set the first degradation batch size in the range $[32, 128]$. In this way, videos in each degradation batch will have inter-prediction in a different pool compared to the previous epoch.

With the frame patches being encoded into a video, they will immediately be decoded back with some extent of quality loss being introduced by the video codec. Note that, in network training, patches are shuffled as a single identity with no correlation to other patches.

\noindent\textbf{Diverse video compression codec standards.} 
Though, based on Section \ref{sec:video_compression}, all video compression standards shared a common pattern, the encoder algorithm designs on how they search for the reference of macroblocks are massively different. To simulate the encoder from a historical perspective, considering both the modern and earlier standards helps the model to learn versatile compression artifacts due to the difference of the video encoder reference searching and architecture designs. This helps us to simulate more real-world video scenarios.
As a result, we choose the most representative and widely used video compression standards: MPEG-2 \cite{mitchell1996mpeg}, MPEG-4 \cite{avaro2000mpeg}, H.264 \cite{schwarz2007overview}, and H.265 \cite{sullivan2012overview}.

\noindent\textbf{Quantization parameter control.}
Quantization Parameter (\textit{QP}) is the direct cause of quality loss in compression. In the real world, people will not explicitly set the \textit{QP} of a video, but it is affected by other settings. For instance, in H.264 and H.265, \textit{QP} is controlled by a Constant Rate Factor (\textit{CRF}) that can be programmed in \textit{ffmpeg}. \textit{CRF} provides an engineered compression rate control for the entire video instead of single macroblocks. The lower the \textit{CRF} is, the less information it would be lost. For MPEG-2 and MPEG-4, we find that controlling bitrate is a better way to manipulate \textit{QP} based on their codec design: to restrict the code size under a certain bitrate, codecs have to increase the \textit{QP} of each macroblock. Hence, we considered \textit{CRF} for H.264 and H.265, and bitrate for MPEG-2 and MPEG-4 in our degradation block implementation.

\noindent\textbf{Encoder speed control.}
Though quantization is widely regarded as the direct source of distortion in compression, and previous works \cite{wu2022animesr, li2021comisr, khani2021efficient, yang2022learned, xiang2021boosting, chan2022investigating} mainly focus on tuning \textit{QP}-related encoder parameters. In Figure \ref{fig:CRF_preset}, our study on video compression standards promotes the finding that encoder speed is a hidden factor that influences \textit{QP} to create video compression artifacts to frames.
To accelerate the video processing speed, video encoders provide various encode speeds (\textit{preset}) from $medium$ (default mode) to $fast$, $faster$, $veryfast$, and even $ultrafast$ speed mode. The faster the encoder speed is, the less time the searching algorithm will have for intra-prediction and inter-prediction. 
Consequently, the predicted reference will be hard to match target macroblocks by pixel comparison, which increases the residual magnitude.

The \textit{QP} factor by itself, therefore, is not enough to encompass real-world video distortions, and the addition of encoder speed control is needed.
In addition, we introduce Frame rate (FPS) control and aspect ratio scaling to augment real-world compression artifact synthesis.

\subsection{Network Architecture and Training}
\label{sec:degradation_model_and_network_architecture}
Given the prevalent success of Transformer-based networks in diverse vision tasks, this work leverages the GRL\cite{li2023efficient} architecture as the foundational model. We enhance this baseline architecture by introducing the video artifact degradation pipeline, thereby enabling the network to learn an extended range of degradation patterns, including those associated with video coding artifacts.

GRL exploits correlations across multiple levels of image hierarchies via Transformer-based architecture, which achieves state-of-the-art performance in generic super-resolution and restoration tasks (\textit{e.g.}, deblurring, JPEG restoration, demosaic, and classical SR). To forge efficient training and to reach a reasonable balance between computation resources and performance, we choose the \textit{small} variation of GRL (Figure \ref{fig:grl_network}) with 3.49M parameters as opposed to the base GRL model that totals 20.2M parameters.

Following Real-ESRGAN, BSRGAN, SwinIR, and GRL\cite{wang2021real, zhang2021designing, liang2021swinir, li2023efficient}, we first train the network with a PSNR-oriented L1 loss. Then we use the trained network parameters to initialize the generator and train a GAN model to boost perceptual quality. The loss function during this phase combines L1 loss, perceptual loss\cite{johnson2016perceptual}, and GAN loss\cite{goodfellow2020generative, ledig2017photo, blau2018the}.

\begin{figure*}[]
    \centering
    \includegraphics[height=0.95\textheight]{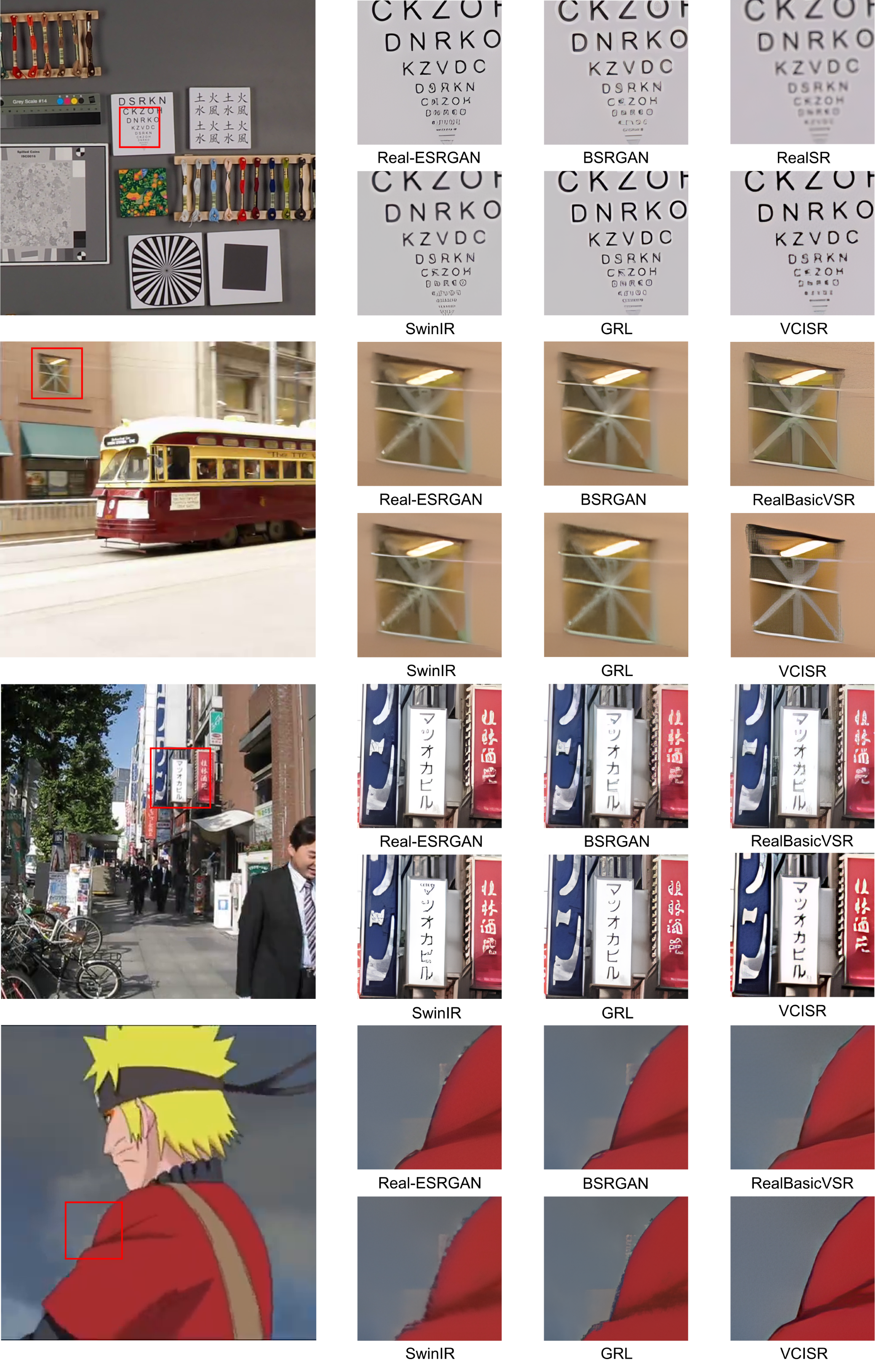}
    \caption{Qualitative comparisons of different methods on $\times 4$ super-resolved images in the DRealSR\cite{wei2020component}, VideoLQ\cite{chan2022investigating}, VideoLQ\cite{chan2022investigating}, and AVC\cite{wu2022animesr} datasets from top to bottom respectively. \textbf{(Zoom in for best view.)}} 
    \label{fig:qualitative}
\end{figure*}

\begin{table*}
\caption{The NIQE\cite{mittal2012making}, BRISQUE\cite{mittal2011blind}, NRQM\cite{ma2017learning}, and CLIP-IQA\cite{wang2023exploring} results of different blind SISR methods on the proposed VC-RealLQ dataset, together with the RealSR\cite{cai2019toward} and DRealSR\cite{wei2020component} dataset. The best and the second best results are remarked in bold font and underlined respectively. We use pyiqa\cite{pyiqa} library to test all datasets.}

\label{tab:ISR_quantitative}
\resizebox{\linewidth}{!}{
\begin{tabular}{c|c|cccc|cccc|cccc}
    \toprule
    
    \multirow{2}{*}{\textbf{Method}} & \multirow{2}{*}{\textbf{\# Params (M)}} & \multicolumn{4}{c|}{\textbf{VC-RealLQ}} & \multicolumn{4}{c|}{\textbf{RealSR-Nikon}} & \multicolumn{4}{c}{\textbf{DRealSR}}\\ \cline{3-14}
     &  & NIQE $\downarrow$ & BRISQUE $\downarrow$ & NRQM $\uparrow$ & CLIPIQA $\uparrow$ & NIQE $\downarrow$ & BRISQUE $\downarrow$ & NRQM $\uparrow$ & CLIPIQA $\uparrow$ & NIQE $\downarrow$ & BRISQUE $\downarrow$ & NRQM $\uparrow$ & CLIPIQA $\uparrow$ \\
     
    \midrule
     RealSR\cite{Ji_2020_CVPR_Workshops} & 16.7 & 5.738 & 37.04 & 5.147 & 0.334 & 7.435 & 57.029 & 3.21  & 0.275 & 8.457 & 56.877 & 3.507 & 0.238 \\
     Real-ESRGAN+\cite{wang2021real} & 16.7 & \underline{4.967} & \underline{29.29} & 5.231 & 0.438 & 4.901 & 31.911 & \underline{5.668} & 0.499 & 4.718 & 29.872 & 5.428 & 0.518 \\
     BSRGAN\cite{zhang2021designing}  & 16.7  & 5.164 & 29.394 & \underline{5.242} & \underline{0.498} & \textbf{4.772} &\textbf{25.382} & \textbf{5.938} & \underline{0.564} & 4.681 & \underline{27.858} & \underline{5.461} & \underline{0.57} \\
     SwinIR\cite{liang2021swinir} & 11.9  & 5.095 & 33.097 & 4.922 & 0.438 & 4.877 & 34.964 & 5.408 & 0.47 & 6.259 & 49.546 & 5.183 & 0.465 \\
     GRL\cite{li2023efficient} & 20.2 & 5.338 & 33.769 & 5.043 & 0.451 & 4.981 & 34.937 & 5.37 & 0.456 & \underline{4.633} & 29.323 & 5.389 & 0.545\\
     VCISR (ours) & \textbf{3.49} & \textbf{4.542} &\textbf{ 16.975} & \textbf{5.479} & \textbf{0.58} & \underline{4.823} & \underline{29.203} & 5.445 & \textbf{0.603} & \textbf{3.983} & \textbf{15.303} & \textbf{5.778} & \textbf{0.646}\\
    \bottomrule
    \end{tabular}
    \vspace{10pt}
}
\end{table*}

\begin{table*}
\caption{The NIQE\cite{mittal2012making}, BRISQUE\cite{mittal2011blind}, NRQM\cite{ma2017learning}, and CLIP-IQA\cite{wang2023exploring} results of different ISR, VSR methods on the REDS\cite{Nah_2019_CVPR_Workshops_REDS}, AVC-RealLQ\cite{wu2022animesr}, and ViedoLQ\cite{chan2022investigating} dataset. The best and the second best results are remarked in bold font and underlined respectively. Due to high computation intensity, following RealBasicVSR\cite{chan2022investigating}, NRQM is computed on the first, middle, and last frames in each sequence. We use pyiqa\cite{pyiqa} library to test all datasets.}
\label{tab:VSR_quantitative}
\resizebox{\linewidth}{!}
{ 
\begin{tabular}{c|c|cccc|cccc|cccc}
    \toprule
    \multirow{2}{*}{\textbf{Method}} & \multirow{2}{*}{\textbf{\# Params (M)}} & \multicolumn{4}{c|}{\textbf{REDS+Blur+MPEG}} & \multicolumn{4}{c|}{\textbf{AVC-RealLQ}} & \multicolumn{4}{c}{\textbf{VideoLQ}}\\ \cline{3-14}
     &  & NIQE $\downarrow$ & BRISQUE $\downarrow$ & NRQM $\uparrow$ & CLIPIQA $\uparrow$ & NIQE $\downarrow$ & BRISQUE $\downarrow$ & NRQM $\uparrow$ & CLIPIQA $\uparrow$ & NIQE $\downarrow$ & BRISQUE $\downarrow$ & NRQM $\uparrow$ & CLIPIQA $\uparrow$ \\
    \midrule
     Real-ESRGAN+\cite{wang2021real} & 16.7 & 5.192 & 32.261 & 5.038 & 0.331 & 8.632 & 36.087 & 5.054 &  0.534 & 4.204 & 29.844 & 5.815 & 0.362 \\
     BSRGAN\cite{zhang2021designing}  & 16.7  & 5.303 & 28.498 & 4.906 & 0.402 & 8.281 & 34.987 & 5.004 & \underline{0.61} & 4.211 & \underline{25.24} & \underline{5.825} & \underline{0.422} \\
     SwinIR\cite{liang2021swinir} & 11.9  & 5.214 & 32.476 & 4.984 & 0.34 & \underline{6.351} & 41.285 & 4.739 & 0.506 & 4.198 & 31.492 & 5.667 & 0.38 \\
     GRL\cite{li2023efficient} & 20.2 & 5.611 & 35.256 & 4.767 & 0.368 & 6.457 & 43.439 & 5.034 & 0.575 & 4.476 & 32.349 & 5.523 & 0.389\\
     RealBasicVSR\cite{chan2022investigating} & 6.3	&\textbf{3.765}	& \underline{14.6} & \textbf{5.43} & \underline{0.41} & 8.639 & \underline{26.013} & \underline{5.07} & 0.583 & \underline{3.766} & 29.03 &\textbf{6.0477}	& 0.376 \\
     DBVSR\cite{pan2021deep} & 25.5 & - & - & - & - & - & - & - & - & 6.7866 & 50.936 &  3.4097 & -\\
    VCISR (ours) & \textbf{3.49} & \underline{4.427} & \textbf{12.709} & \underline{5.207} & \textbf{0.508} & \textbf{5.515} & \textbf{19.529} & \textbf{5.236} & \textbf{0.689} & \textbf{3.725} & \textbf{19.372} &  5.735 & \textbf{0.488}\\
    \bottomrule
    \end{tabular}
}
\end{table*}

\section{Experiments}
\subsection{Experimental Setup}
\noindent\textbf{Datasets.}
We train the proposed model with the DIV2K dataset\cite{agustsson2017ntire}, which has gained prominence in the field of image super-resolution. To validate the efficacy of the presented approach, we perform comprehensive evaluations on both image super-resolution and video super-resolution tasks. Specifically, for image super-resolution, we employ the RealSR-Nikon\cite{cai2019toward} and the DRealSR\cite{wei2020component} dataset, capitalizing on their distinct characteristics. Additionally, to enhance the diversity of degradation patterns in our test set, we bring in a specialized dataset simulating video coding-introduced degradations. More details are elaborated in Section \ref{sec:vcreal_lq}.
Furthermore, to assess the versatility of our proposed method, we subjected it to rigorous testing on widely used Video Super-Resolution (VSR) datasets, including the REDS\cite{Nah_2019_CVPR_Workshops_REDS}, AVC-RealLQ\cite{wu2022animesr}, and VideoLQ\cite{chan2022investigating}. These datasets are chosen for their relevance and ability to provide insights into the method's performance across varying video content and quality levels. Through this meticulous evaluation process, we aim to showcase the robustness and adaptability of our approach across both image and video super-resolution domains.

\noindent\textbf{Training details.}
The neural network training is performed on one Nvidia RTX 4090 GPU. For the first stage, we train the network with L1 loss for 700\textit{K} iterations, employing a batch size of 12. The Adam optimizer\cite{kingma2014adam} is adopted with a learning rate of $2\times 10^{-4}$, which is decayed by half every 50\textit{K} iterations. 
In the subsequent adversarial training stage, the model is trained for 280\textit{K} iterations, employing a batch size of 6. The learning rate for this stage is set at $1\times 10^{-4}$ and decayed to half every 20\textit{K} iterations. The adversarial training of our model adhered to a weighted perceptual loss \cite{johnson2016perceptual} and employed a U-Net discriminator with spectral normalization\cite{miyato2018spectral}. 
The perceptual loss utilized the same pre-trained VGG-19 \cite{johnson2016perceptual}, with weight coefficients $\{0.1, 0.1, 1, 1, 1\}$ corresponding to feature maps $\{\mathtt{conv1}, ... \mathtt{conv5}\}$.

In accordance with the details outlined in the previous Section \ref{sec:video_artifact_degradation_block}, a preprocessing step is performed before commencing training. Large images are initially cropped into non-overlapping high-resolution (HR) patches with a resolution of $360\times360$. Consequently, for our $\times 4$ scaling task, the corresponding low-resolution (LR) patch size is established at $90\times90$, constituting the output size for all degradation models. This choice of patch size is informed by the characteristics of H.265 \cite{sullivan2012overview}, where the basic processing unit reaches dimensions of $64\times64$. This selection is based on the belief that a relatively larger image size is imperative for optimizing the effectiveness of the video encoder during the intra-prediction and inter-prediction stages. In total, our training dataset is expended to 12,630 HR patches as input during the training phase, all sourced from 800 DIV2K training images. 

\noindent\textbf{Degradation details.}
The proposed degradation model runs in every epoch to prepare the LR, HR pair used for training. The degradation batch size is set to 128. We referred to \cite{wang2021real} for degradation settings in noise, blur, resize, and JPEG compression blocks.
Our video compression degradation block uses standard codecs from MPEG-2, MPEG-4, H.264, and H.265 with probability $[0.2, 0.2, 0.4, 0.2]$. H.264 and H.265 control quantization through CRF with range $[20, 32]$ and $[25, 37]$ respectively. 
MPEG-2 and MPEG-4 control quantization through bitrate restriction in the range $[4000, 6000]$ Kbit/s. 
For all standard codecs, encoder speed setting \textit{preset} is chosen from $\{slow, medium, fast, faster, superfast\}$ with probability $\{0.1, 0.5 , 0.25,0.12, 0.03\}$ respectively. For aspect ratio scaling, the width scales in $[0.85, 1.35]$ with a probability of $\{0.2, 0.4, 0.4\}$ to shrink, expand, or keep the same. FPS is chosen from $[16, 30]$. Since the parameter range that MPEG-2 can support is limited, we set it to a fixed 25 FPS without any aspect ratio scaling.
Other video compression parameters for all standards are fixed. For example, pixel encode format is uniformly set to be \textit{YUV420p}. We leave the rest compression parameters to be decided by codec, like $Profile$ and $Levels$.

\subsection{Results and Analysis}
We evaluate our proposed VCISR qualitatively and quantitatively in comparison with several state-of-the-art approaches, including RealSR \cite{Ji_2020_CVPR_Workshops}, Real-ESRGAN+\cite{wang2021real}, BSRGAN\cite{zhang2021designing}, SwinIR\cite{liang2021swinir}, and GRL\cite{li2023efficient}. For the VSR task, results from RealBasicVSR\cite{chan2022investigating} and DBVSR\cite{pan2021deep} are involved additionally.

\noindent\textbf{Qualitative analysis.} Figure \ref{fig:qualitative} captures a couple of illustrative samples from a collection of image and video test datasets with diverse scenes and contents. Our work is capable of restoring textual information without introducing ringing artifacts and unfaithful reconstructions, which owes to a more complex degradation model as we proposed in VCISR. For scenes captured under high motion, our method demonstrates an even better ability to resolve blurry artifacts compared to the VSR method. As for the Anime content, the HR image super-resolved by VCISR does not include block artifacts and fringes. Both the edge and color details are better restored by our scheme.

\noindent\textbf{Quantitative analysis}. For quantitative comparisons, since not all datasets have paired HR ground truth, following\cite{wu2022animesr, li2023efficient, wang2021real, zhang2021designing, chan2022investigating}, we adopt no-reference quality assessments, instead of PSNR or SSIM, to evaluate the resulting HR images. Table \ref{tab:ISR_quantitative} and \ref{tab:VSR_quantitative} presents our quantitative measurements based on the no-reference metrics that are widely used in previous real-world image and video SR works\cite{wang2021real, chan2022investigating, ji2020real}: NIQE\cite{mittal2012making}, BRISQUE\cite{mittal2011blind}, and NRQM\cite{ma2017learning}. In addition, we employ a SOTA learning-based CLIP-IQA metric\cite{wang2023exploring}, which gives a score closer to human perceptual assessment. 

The proposed method exhibits similar or better quantitative results on the tested data for blind SISR compared to recently published works with much fewer network parameters. For the VSR task, we showcase that involving a video-based degradation module can produce effective restorations even without leveraging the temporal correlations by referring to previous frames, which typically require more memory and compute complexity during inference.

\subsection{VC-RealLQ Dataset}
\label{sec:vcreal_lq}
Most of the existing image datasets \cite{wei2020component, ignatov2017dslr, cai2019toward} only consider photography as the input source. This makes them dismiss compression artifacts in the real world. For the convenience of future researchers, we propose an image-based dataset that contains 35 images each with versatile compression artifacts (\eg ringing artifacts, mosquito noise, blockiness, and color bleeding). They come from video sequence screenshots after H.264 \cite{schwarz2007overview}, H.265 \cite{sullivan2012overview}, or WebP\cite{si2016research} compression on video datasets\cite{wu2022animesr, chan2022investigating, mercat2020uvg, sullivan2012overview}, and unknown compression degradation from online resources. We select these images with different resolutions, contents, and different levels of degradation.

\section{Conclusion}
This work presents a novel and versatile video-codec-based degradation module to enrich the existing image SR degradation pipelines and improve its proximity to practical exercises. The method we propose for LR image synthesis can simulate distortions from a variety of video codecs without the presence of video data, which leads to \textbf{1)} a generic element that could be added to existing SR degradation pipelines; \textbf{2)} better-approximated artifacts to cover more complicated quality loss in real-world images and videos. We demonstrate the viability of this approach by evaluating a unified network trained with the proposed degradation flow on real-world ISR and VSR datasets. In both tasks, our work exhibits similar or better performance compared to other state-of-the-art methods with fewer network parameters and compute complexity.

\section{Acknowledgements}
We thank Xiaotong Chen, Huijie Zhang, and Jimmy Tobin for feedback on drafts and early discussions. 

{\small
\bibliographystyle{ieee_fullname}
\bibliography{egbib}

\begin{thebibliography}{10}\itemsep=-1pt

\bibitem{agustsson2017ntire}
Eirikur Agustsson and Radu Timofte.
\newblock Ntire 2017 challenge on single image super-resolution: Dataset and study.
\newblock In {\em Proceedings of the IEEE conference on computer vision and pattern recognition workshops}, pages 126--135, 2017.

\bibitem{ahmed1974discrete}
Nasir Ahmed, T\_ Natarajan, and Kamisetty~R Rao.
\newblock Discrete cosine transform.
\newblock {\em IEEE transactions on Computers}, 100(1):90--93, 1974.

\bibitem{avaro2000mpeg}
Olivier Avaro, Alexandros Eleftheriadis, Carsten Herpel, Ganesh Rajan, and Liam Ward.
\newblock Mpeg-4 systems: overview.
\newblock {\em Signal Processing: Image Communication}, 15(4-5):281--298, 2000.

\bibitem{blau2018the}
Yochai Blau and Tomer Michaeli.
\newblock The perception-distortion tradeoff.
\newblock In {\em 2018 IEEE/CVF Conference on Computer Vision and Pattern Recognition}, pages 6228--6237, 2018.

\bibitem{caballero2017real}
Jose Caballero, Christian Ledig, Andrew Aitken, Alejandro Acosta, Johannes Totz, Zehan Wang, and Wenzhe Shi.
\newblock Real-time video super-resolution with spatio-temporal networks and motion compensation.
\newblock In {\em Proceedings of the IEEE conference on computer vision and pattern recognition}, pages 4778--4787, 2017.

\bibitem{cai2019toward}
Jianrui Cai, Hui Zeng, Hongwei Yong, Zisheng Cao, and Lei Zhang.
\newblock Toward real-world single image super-resolution: A new benchmark and a new model.
\newblock In {\em Proceedings of the IEEE/CVF International Conference on Computer Vision}, pages 3086--3095, 2019.

\bibitem{chan2022investigating}
Kelvin~CK Chan, Shangchen Zhou, Xiangyu Xu, and Chen~Change Loy.
\newblock Investigating tradeoffs in real-world video super-resolution.
\newblock In {\em Proceedings of the IEEE/CVF Conference on Computer Vision and Pattern Recognition}, pages 5962--5971, 2022.

\bibitem{pyiqa}
Chaofeng Chen and Jiadi Mo.
\newblock {IQA-PyTorch}: Pytorch toolbox for image quality assessment.
\newblock [Online]. Available: \url{https://github.com/chaofengc/IQA-PyTorch}, 2022.

\bibitem{chen2023search}
Yu Chen, Mingyu Yang, and Hun-Seok Kim.
\newblock Search for efficient deep visual-inertial odometry through neural architecture search.
\newblock In {\em ICASSP 2023-2023 IEEE International Conference on Acoustics, Speech and Signal Processing (ICASSP)}, pages 1--5. IEEE, 2023.

\bibitem{dong2015image}
Chao Dong, Chen~Change Loy, Kaiming He, and Xiaoou Tang.
\newblock Image super-resolution using deep convolutional networks.
\newblock {\em IEEE transactions on pattern analysis and machine intelligence}, 38(2):295--307, 2015.

\bibitem{dosovitskiy2020image}
Alexey Dosovitskiy, Lucas Beyer, Alexander Kolesnikov, Dirk Weissenborn, Xiaohua Zhai, Thomas Unterthiner, Mostafa Dehghani, Matthias Minderer, Georg Heigold, Sylvain Gelly, et~al.
\newblock An image is worth 16x16 words: Transformers for image recognition at scale.
\newblock {\em arXiv preprint arXiv:2010.11929}, 2020.

\bibitem{efrat2013accurate}
Netalee Efrat, Daniel Glasner, Alexander Apartsin, Boaz Nadler, and Anat Levin.
\newblock Accurate blur models vs. image priors in single image super-resolution.
\newblock In {\em Proceedings of the IEEE International Conference on Computer Vision}, pages 2832--2839, 2013.

\bibitem{fu2022kxnet}
Jiahong Fu, Hong Wang, Qi Xie, Qian Zhao, Deyu Meng, and Zongben Xu.
\newblock Kxnet: A model-driven deep neural network for blind super-resolution.
\newblock In {\em Computer Vision--ECCV 2022: 17th European Conference, Tel Aviv, Israel, October 23--27, 2022, Proceedings, Part XIX}, pages 235--253. Springer, 2022.

\bibitem{goodfellow2020generative}
Ian Goodfellow, Jean Pouget-Abadie, Mehdi Mirza, Bing Xu, David Warde-Farley, Sherjil Ozair, Aaron Courville, and Yoshua Bengio.
\newblock Generative adversarial networks.
\newblock {\em Communications of the ACM}, 63(11):139--144, 2020.

\bibitem{He2015DeepRL}
Kaiming He, X. Zhang, Shaoqing Ren, and Jian Sun.
\newblock Deep residual learning for image recognition.
\newblock {\em 2016 IEEE Conference on Computer Vision and Pattern Recognition (CVPR)}, pages 770--778, 2015.

\bibitem{ignatov2017dslr}
Andrey Ignatov, Nikolay Kobyshev, Radu Timofte, Kenneth Vanhoey, and Luc Van~Gool.
\newblock Dslr-quality photos on mobile devices with deep convolutional networks.
\newblock In {\em Proceedings of the IEEE international conference on computer vision}, pages 3277--3285, 2017.

\bibitem{ji2020real}
Xiaozhong Ji, Yun Cao, Ying Tai, Chengjie Wang, Jilin Li, and Feiyue Huang.
\newblock Real-world super-resolution via kernel estimation and noise injection.
\newblock In {\em proceedings of the IEEE/CVF conference on computer vision and pattern recognition workshops}, pages 466--467, 2020.

\bibitem{Ji_2020_CVPR_Workshops}
Xiaozhong Ji, Yun Cao, Ying Tai, Chengjie Wang, Jilin Li, and Feiyue Huang.
\newblock Real-world super-resolution via kernel estimation and noise injection.
\newblock In {\em The IEEE/CVF Conference on Computer Vision and Pattern Recognition (CVPR) Workshops}, June 2020.

\bibitem{johnson2016perceptual}
Justin Johnson, Alexandre Alahi, and Li Fei-Fei.
\newblock Perceptual losses for real-time style transfer and super-resolution.
\newblock In {\em Computer Vision--ECCV 2016: 14th European Conference, Amsterdam, The Netherlands, October 11-14, 2016, Proceedings, Part II 14}, pages 694--711. Springer, 2016.

\bibitem{khani2021efficient}
Mehrdad Khani, Vibhaalakshmi Sivaraman, and Mohammad Alizadeh.
\newblock Efficient video compression via content-adaptive super-resolution.
\newblock In {\em Proceedings of the IEEE/CVF International Conference on Computer Vision}, pages 4521--4530, 2021.

\bibitem{kingma2014adam}
Diederik~P Kingma and Jimmy Ba.
\newblock Adam: A method for stochastic optimization.
\newblock {\em arXiv preprint arXiv:1412.6980}, 2014.

\bibitem{ledig2017photo}
Christian Ledig, Lucas Theis, Ferenc Husz{\'a}r, Jose Caballero, Andrew Cunningham, Alejandro Acosta, Andrew Aitken, Alykhan Tejani, Johannes Totz, Zehan Wang, et~al.
\newblock Photo-realistic single image super-resolution using a generative adversarial network.
\newblock In {\em Proceedings of the IEEE conference on computer vision and pattern recognition}, pages 4681--4690, 2017.

\bibitem{li2023interpretable}
Gaotang Li, Marlena Duda, Xiang Zhang, Danai Koutra, and Yujun Yan.
\newblock Interpretable sparsification of brain graphs: Better practices and effective designs for graph neural networks.
\newblock {\em arXiv preprint arXiv:2306.14375}, 2023.

\bibitem{li2023efficient}
Yawei Li, Yuchen Fan, Xiaoyu Xiang, Denis Demandolx, Rakesh Ranjan, Radu Timofte, and Luc Van~Gool.
\newblock Efficient and explicit modelling of image hierarchies for image restoration.
\newblock In {\em Proceedings of the IEEE/CVF Conference on Computer Vision and Pattern Recognition}, pages 18278--18289, 2023.

\bibitem{li2021comisr}
Yinxiao Li, Pengchong Jin, Feng Yang, Ce Liu, Ming-Hsuan Yang, and Peyman Milanfar.
\newblock Comisr: Compression-informed video super-resolution.
\newblock In {\em Proceedings of the IEEE/CVF International Conference on Computer Vision}, pages 2543--2552, 2021.

\bibitem{liang2022vrt}
Jingyun Liang, Jiezhang Cao, Yuchen Fan, Kai Zhang, Rakesh Ranjan, Yawei Li, Radu Timofte, and Luc Van~Gool.
\newblock Vrt: A video restoration transformer.
\newblock {\em arXiv preprint arXiv:2201.12288}, 2022.

\bibitem{liang2021swinir}
Jingyun Liang, Jiezhang Cao, Guolei Sun, Kai Zhang, Luc Van~Gool, and Radu Timofte.
\newblock Swinir: Image restoration using swin transformer.
\newblock In {\em Proceedings of the IEEE/CVF international conference on computer vision}, pages 1833--1844, 2021.

\bibitem{liang2021mutual}
Jingyun Liang, Guolei Sun, Kai Zhang, Luc Van~Gool, and Radu Timofte.
\newblock Mutual affine network for spatially variant kernel estimation in blind image super-resolution.
\newblock In {\em Proceedings of the IEEE/CVF International Conference on Computer Vision}, pages 4096--4105, 2021.

\bibitem{lim2017enhanced}
Bee Lim, Sanghyun Son, Heewon Kim, Seungjun Nah, and Kyoung Mu~Lee.
\newblock Enhanced deep residual networks for single image super-resolution.
\newblock In {\em Proceedings of the IEEE conference on computer vision and pattern recognition workshops}, pages 136--144, 2017.

\bibitem{liu2022blind}
Anran Liu, Yihao Liu, Jinjin Gu, Yu Qiao, and Chao Dong.
\newblock Blind image super-resolution: A survey and beyond.
\newblock {\em IEEE Transactions on Pattern Analysis and Machine Intelligence}, 2022.

\bibitem{liu2020unified}
Bowen Liu, Ang Cao, and Hun-Seok Kim.
\newblock Unified signal compression using generative adversarial networks.
\newblock In {\em ICASSP 2020-2020 IEEE International Conference on Acoustics, Speech and Signal Processing (ICASSP)}, pages 3177--3181. IEEE, 2020.

\bibitem{liu2021deep}
Bowen Liu, Yu Chen, Shiyu Liu, and Hun-Seok Kim.
\newblock Deep learning in latent space for video prediction and compression.
\newblock In {\em Proceedings of the IEEE/CVF conference on computer vision and pattern recognition}, pages 701--710, 2021.

\bibitem{liu2023mmvc}
Bowen Liu, Yu Chen, Rakesh~Chowdary Machineni, Shiyu Liu, and Hun-Seok Kim.
\newblock Mmvc: Learned multi-mode video compression with block-based prediction mode selection and density-adaptive entropy coding.
\newblock In {\em Proceedings of the IEEE/CVF Conference on Computer Vision and Pattern Recognition}, pages 18487--18496, 2023.

\bibitem{liu2013bayesian}
Ce Liu and Deqing Sun.
\newblock On bayesian adaptive video super resolution.
\newblock {\em IEEE transactions on pattern analysis and machine intelligence}, 36(2):346--360, 2013.

\bibitem{liu2020estimating}
Yu-Qi Liu, Xin Du, Hui-Liang Shen, and Shu-Jie Chen.
\newblock Estimating generalized gaussian blur kernels for out-of-focus image deblurring.
\newblock {\em IEEE Transactions on circuits and systems for video technology}, 31(3):829--843, 2020.

\bibitem{luo2022deep}
Ziwei Luo, Haibin Huang, Lei Yu, Youwei Li, Haoqiang Fan, and Shuaicheng Liu.
\newblock Deep constrained least squares for blind image super-resolution.
\newblock In {\em Proceedings of the IEEE/CVF Conference on Computer Vision and Pattern Recognition}, pages 17642--17652, 2022.

\bibitem{luo2022learning}
Zhengxiong Luo, Yan Huang, Shang Li, Liang Wang, and Tieniu Tan.
\newblock Learning the degradation distribution for blind image super-resolution.
\newblock In {\em Proceedings of the IEEE/CVF Conference on Computer Vision and Pattern Recognition}, pages 6063--6072, 2022.

\bibitem{MaQBL23}
Chenyang Ma, Xinchi Qiu, Daniel~J. Beutel, and Nicholas~D. Lane.
\newblock Gradient-less federated gradient boosting tree with learnable learning rates.
\newblock In {\em Proceedings of the 3rd Workshop on Machine Learning and Systems, EuroMLSys 2023, Rome, Italy, 8 May 2023}, pages 56--63. {ACM}, 2023.

\bibitem{ma2017learning}
Chao Ma, Chih-Yuan Yang, Xiaokang Yang, and Ming-Hsuan Yang.
\newblock Learning a no-reference quality metric for single-image super-resolution.
\newblock {\em Computer Vision and Image Understanding}, 158:1--16, 2017.

\bibitem{mercat2020uvg}
Alexandre Mercat, Marko Viitanen, and Jarno Vanne.
\newblock Uvg dataset: 50/120fps 4k sequences for video codec analysis and development.
\newblock In {\em Proceedings of the 11th ACM Multimedia Systems Conference}, pages 297--302, 2020.

\bibitem{mitchell1996mpeg}
Joan~L Mitchell, William~B Pennebaker, Chad~E Fogg, Didier~J LeGall, Joan~L Mitchell, William~B Pennebaker, Chad~E Fogg, and Didier~J LeGall.
\newblock Mpeg-2 overview.
\newblock {\em MPEG Video Compression Standard}, pages 171--186, 1996.

\bibitem{mittal2011blind}
Anish Mittal, Anush~K Moorthy, and Alan~C Bovik.
\newblock Blind/referenceless image spatial quality evaluator.
\newblock In {\em 2011 conference record of the forty fifth asilomar conference on signals, systems and computers (ASILOMAR)}, pages 723--727. IEEE, 2011.

\bibitem{mittal2012making}
Anish Mittal, Rajiv Soundararajan, and Alan~C Bovik.
\newblock Making a “completely blind” image quality analyzer.
\newblock {\em IEEE Signal processing letters}, 20(3):209--212, 2012.

\bibitem{miyato2018spectral}
Takeru Miyato, Toshiki Kataoka, Masanori Koyama, and Yuichi Yoshida.
\newblock Spectral normalization for generative adversarial networks.
\newblock {\em arXiv preprint arXiv:1802.05957}, 2018.

\bibitem{Nah_2019_CVPR_Workshops_REDS}
Seungjun Nah, Sungyong Baik, Seokil Hong, Gyeongsik Moon, Sanghyun Son, Radu Timofte, and Kyoung~Mu Lee.
\newblock Ntire 2019 challenge on video deblurring and super-resolution: Dataset and study.
\newblock In {\em CVPR Workshops}, June 2019.

\bibitem{pan2021deep}
Jinshan Pan, Haoran Bai, Jiangxin Dong, Jiawei Zhang, and Jinhui Tang.
\newblock Deep blind video super-resolution.
\newblock In {\em Proceedings of the IEEE/CVF International Conference on Computer Vision}, pages 4811--4820, 2021.

\bibitem{schwarz2007overview}
Heiko Schwarz, Detlev Marpe, and Thomas Wiegand.
\newblock Overview of the scalable video coding extension of the h. 264/avc standard.
\newblock {\em IEEE Transactions on circuits and systems for video technology}, 17(9):1103--1120, 2007.

\bibitem{shi2016real}
Wenzhe Shi, Jose Caballero, Ferenc Husz{\'a}r, Johannes Totz, Andrew~P Aitken, Rob Bishop, Daniel Rueckert, and Zehan Wang.
\newblock Real-time single image and video super-resolution using an efficient sub-pixel convolutional neural network.
\newblock In {\em Proceedings of the IEEE conference on computer vision and pattern recognition}, pages 1874--1883, 2016.

\bibitem{si2016research}
Zhanjun Si and Ke Shen.
\newblock Research on the webp image format.
\newblock In {\em Advanced graphic communications, packaging technology and materials}, pages 271--277. Springer, 2016.

\bibitem{sullivan2012overview}
Gary~J Sullivan, Jens-Rainer Ohm, Woo-Jin Han, and Thomas Wiegand.
\newblock Overview of the high efficiency video coding (hevc) standard.
\newblock {\em IEEE Transactions on circuits and systems for video technology}, 22(12):1649--1668, 2012.

\bibitem{wallace1992jpeg}
Gregory~K Wallace.
\newblock The jpeg still picture compression standard.
\newblock {\em IEEE transactions on consumer electronics}, 38(1):xviii--xxxiv, 1992.

\bibitem{wang2023exploring}
Jianyi Wang, Kelvin~CK Chan, and Chen~Change Loy.
\newblock Exploring clip for assessing the look and feel of images.
\newblock In {\em Proceedings of the AAAI Conference on Artificial Intelligence}, volume~37, pages 2555--2563, 2023.

\bibitem{wang2023exploiting}
Jianyi Wang, Zongsheng Yue, Shangchen Zhou, Kelvin~CK Chan, and Chen~Change Loy.
\newblock Exploiting diffusion prior for real-world image super-resolution.
\newblock {\em arXiv preprint arXiv:2305.07015}, 2023.

\bibitem{wang2021real}
Xintao Wang, Liangbin Xie, Chao Dong, and Ying Shan.
\newblock Real-esrgan: Training real-world blind super-resolution with pure synthetic data.
\newblock In {\em Proceedings of the IEEE/CVF international conference on computer vision}, pages 1905--1914, 2021.

\bibitem{wang2018esrgan}
Xintao Wang, Ke Yu, Shixiang Wu, Jinjin Gu, Yihao Liu, Chao Dong, Yu Qiao, and Chen Change~Loy.
\newblock Esrgan: Enhanced super-resolution generative adversarial networks.
\newblock In {\em Proceedings of the European conference on computer vision (ECCV) workshops}, pages 0--0, 2018.

\bibitem{wei2020component}
Pengxu Wei, Ziwei Xie, Hannan Lu, Zongyuan Zhan, Qixiang Ye, Wangmeng Zuo, and Liang Lin.
\newblock Component divide-and-conquer for real-world image super-resolution.
\newblock In {\em Computer Vision--ECCV 2020: 16th European Conference, Glasgow, UK, August 23--28, 2020, Proceedings, Part VIII 16}, pages 101--117. Springer, 2020.

\bibitem{Wu_2023_boosting}
Shaokai Wu and Fengyu Yang.
\newblock Boosting detection in crowd analysis via underutilized output features.
\newblock In {\em Proceedings of the IEEE/CVF Conference on Computer Vision and Pattern Recognition (CVPR)}, pages 15609--15618, June 2023.

\bibitem{wu2022adma}
Xintian Wu, Hanbin Zhao, Liangli Zheng, Shouhong Ding, and Xi Li.
\newblock Adma-gan: Attribute-driven memory augmented gans for text-to-image generation.
\newblock In {\em Proceedings of the 30th ACM International Conference on Multimedia}, pages 1593--1602, 2022.

\bibitem{wu2022animesr}
Yanze Wu, Xintao Wang, Gen Li, and Ying Shan.
\newblock Animesr: Learning real-world super-resolution models for animation videos.
\newblock {\em arXiv preprint arXiv:2206.07038}, 2022.

\bibitem{xiang2021boosting}
Xiaoyu Xiang, Qian Lin, and Jan~P Allebach.
\newblock Boosting high-level vision with joint compression artifacts reduction and super-resolution.
\newblock In {\em 2020 25th International Conference on Pattern Recognition (ICPR)}, pages 2390--2397. IEEE, 2021.

\bibitem{Yang2022SparseAC}
Fengyu Yang and Chenyan Ma.
\newblock Sparse and complete latent organization for geospatial semantic segmentation.
\newblock {\em 2022 IEEE/CVF Conference on Computer Vision and Pattern Recognition (CVPR)}, pages 1799--1808, 2022.

\bibitem{yang2022touch}
Fengyu Yang, Chenyang Ma, Jiacheng Zhang, Jing Zhu, Wenzhen Yuan, and Andrew Owens.
\newblock Touch and go: Learning from human-collected vision and touch.
\newblock {\em Neural Information Processing Systems (NeurIPS) - Datasets and Benchmarks Track}, 2022.

\bibitem{yang2023generating}
Fengyu Yang, Jiacheng Zhang, and Andrew Owens.
\newblock Generating visual scenes from touch.
\newblock {\em International Conference on Computer Vision (ICCV)}, 2023.

\bibitem{yang2022learned}
Jiayu Yang, Chunhui Yang, Fei Xiong, Feng Wang, and Ronggang Wang.
\newblock Learned low bitrate video compression with space-time super-resolution.
\newblock In {\em Proceedings of the IEEE/CVF Conference on Computer Vision and Pattern Recognition}, pages 1786--1790, 2022.

\bibitem{yang2022ntire}
Ren Yang, Radu Timofte, Meisong Zheng, Qunliang Xing, Minglang Qiao, Mai Xu, Lai Jiang, Huaida Liu, Ying Chen, Youcheng Ben, et~al.
\newblock Ntire 2022 challenge on super-resolution and quality enhancement of compressed video: Dataset, methods and results.
\newblock In {\em Proceedings of the IEEE/CVF Conference on Computer Vision and Pattern Recognition}, pages 1221--1238, 2022.

\bibitem{yue2022blind}
Zongsheng Yue, Qian Zhao, Jianwen Xie, Lei Zhang, Deyu Meng, and Kwan-Yee~K Wong.
\newblock Blind image super-resolution with elaborate degradation modeling on noise and kernel.
\newblock In {\em Proceedings of the IEEE/CVF Conference on Computer Vision and Pattern Recognition}, pages 2128--2138, 2022.

\bibitem{zhang2023emergence}
Huijie Zhang, Jinfan Zhou, Yifu Lu, Minzhe Guo, Liyue Shen, and Qing Qu.
\newblock The emergence of reproducibility and consistency in diffusion models.
\newblock {\em arXiv preprint arXiv:2310.05264}, 2023.

\bibitem{zhang2021designing}
Kai Zhang, Jingyun Liang, Luc Van~Gool, and Radu Timofte.
\newblock Designing a practical degradation model for deep blind image super-resolution.
\newblock In {\em Proceedings of the IEEE/CVF International Conference on Computer Vision}, pages 4791--4800, 2021.

\bibitem{zhang2022deep}
Kaihao Zhang, Wenqi Ren, Wenhan Luo, Wei-Sheng Lai, Bj{\"o}rn Stenger, Ming-Hsuan Yang, and Hongdong Li.
\newblock Deep image deblurring: A survey.
\newblock {\em International Journal of Computer Vision}, 130(9):2103--2130, 2022.

\bibitem{zhang2022closer}
Wenlong Zhang, Guangyuan Shi, Yihao Liu, Chao Dong, and Xiao-Ming Wu.
\newblock A closer look at blind super-resolution: Degradation models, baselines, and performance upper bounds.
\newblock In {\em Proceedings of the IEEE/CVF Conference on Computer Vision and Pattern Recognition}, pages 527--536, 2022.

\bibitem{Zhao2022RBCRT}
Hanbin Zhao, Fengyu Yang, Xinghe Fu, and Xi Li.
\newblock Rbc: Rectifying the biased context in continual semantic segmentation.
\newblock {\em ArXiv}, abs/2203.08404, 2022.

\bibitem{zhao2023quality}
Kai Zhao, Kun Yuan, Ming Sun, Mading Li, and Xing Wen.
\newblock Quality-aware pre-trained models for blind image quality assessment.
\newblock In {\em Proceedings of the IEEE/CVF Conference on Computer Vision and Pattern Recognition}, pages 22302--22313, 2023.

\end{thebibliography}
}

\end{document}